\documentstyle[preprint,aps]{revtex}
\begin{document}
\draft
\title{Numerical simulations of scattering speckle from phase ordering systems}
\author{Gregory Brown$^1$ \and Per Arne Rikvold$^{1,2,3}$ 
        \and Martin Grant$^1$ }
\address{$^1$Centre for the Physics of Materials, McGill University,\\
	     3600 rue University, Montr{\'{e}}al, Qu{\'{e}}bec, Canada H3A 2T8\\
         $^2$Department of Fundamental Sciences, 
	     Faculty of Integrated Human Studies\\
	     Kyoto University, Kyoto 606, Japan\\
         $^3$Center for Materials Research and Technology,\\
             Supercomputer Computations Research Institute,   
             and Department of Physics\\
             Florida State University, Tallahassee, Florida 32306-3016\\
}

\date{\today}
\maketitle
\begin{abstract}

The scattering of coherent X-rays from dynamically evolving systems is
currently becoming experimentally feasible. The scattered beam
produces a pattern of bright and dark speckles, which fluctuate almost
independently in time and can be used to study the dynamics of the
system. Here we report large-scale computer simulations of the speckle
dynamics for a phase ordering system, using a two-dimensional model
quenched through an order--disorder transition into the two-phase
regime.  The intensity at each wave vector ${\bf k}$ is found to be an
exponentially distributed random variable.  The scaling hypothesis is
extended to the two-time correlation function of the scattering
intensity at a given wave vector, ${\rm Corr}({\bf k};t_1,t_2)$. The
characteristic decay time difference for the correlation function, 
$|t_1 \! - \! t_2|_c$, is found to scale as $k|t_1 \! + \! t_2|^{1/2}$.

\end{abstract}

\newpage

\section{Introduction} 
\label{sec:Int}

When electromagnetic radiation is scattered from an inhomogeneous
material, it undergoes phase changes which depend on which part of the
material it scatters from. The resulting pattern of constructive and
destructive interference at the detector depends sensitively on the
geometry of domains in the scattering region. If the incident
radiation is not coherent over the scattering region, this dependence
is washed out by the phase differences in the incident beam.  However, if the
incident beam can be made coherent over the scattering volume, a
speckled scattering pattern such as that shown in Fig.~1 is
observed. The intensity fluctuations associated with the speckle
pattern are the basis of photon correlation experiments. Over the last
three decades, lasers have become available as sources of coherent
radiation at wavelengths ranging from the ultraviolet to the infrared
\cite{Chu}.

The much shorter wavelength of X-rays allows materials to be studied
at smaller length scales than is possible with light, and the greater
penetration allows the study of optically opaque materials.  However,
coherent X-rays have only become readily available with the emergence
of high-intensity synchrotron photon sources. Recently, speckle
patterns have been observed in several experiments using coherent
synchrotron X-rays.  For example, diffusion rates in gold colloids
were determined from the time Brownian motion requires to change the
X-ray speckle pattern \cite{Dierker:95,Chu:95}, and speckle from
coherent X-rays has also been used to study equilibrium fluctuations
in Fe$_3$Al near an order-disorder transition \cite{Brauer:95}.

An important situation for analyzing phase ordering processes is when
the system is characterized by a single length scale $R$,
corresponding to the average domain size, that grows asymptotically
with time $t$ as $R \propto t^n$, where $n$ is known as the growth
exponent.  If the dynamics are controlled by diffusion, $n=1/2$ for a
non-conserved order parameter and $n=1/3$ for a conserved order
parameter.  The significance of this for scattering experiments is
that the scattering intensity at wavevector ${\bf k}$ is proportional
to the structure factor $S({\bf k},t)$. At all times in the asymptotic
scaling regime the average structure factor, $\langle S(k,t)\rangle$,
can be expressed in terms of a scaling function which depends on $k$
and $t$ only through a dimensionless wavevector $q \propto k t^n$:
\begin{equation}
t^{-dn} \langle S(k,t) \rangle \propto \langle I(q)\rangle \;, 
\label{eq:scalq}
\end{equation}
where $d$ is the spatial dimensionality.  

An alternative way to
consider this scaling behavior, which is more useful in describing the
time dependence of the fluctuations in the scattering intensities at
different wavevectors, is the following.  One observes that the
dynamics of structures on different length scales $k^{-1}$ become
statistically indistinguishable when viewed on the corresponding time
scales $\theta \propto tk^{1/n} \equiv q^{1/n}$.  This can be
expressed in a scaling relation equivalent to Eq.~(\ref{eq:scalq}):
\begin{equation}
k^d \langle S(k,t)\rangle  \propto \langle F(\theta)\rangle \equiv 
\theta^{nd} \langle I(\theta^n)\rangle \;. 
\label{eq:scalt}
\end{equation}

In this paper we present a numerical study of a simple model for the
scattering speckle observed in coherent X-ray scattering from systems
undergoing phase ordering following a quench through a second-order
order--disorder phase transition.  This is the standard time-dependent
Ginzburg-Landau (TDGL) equation with non-conserved order parameter,
equivalent to Model A in the classification scheme of Hohenberg and
Halperin \cite{Hohenberg:77}.  Using standard Langevin dynamics, we
study the fluctuations in $\bf k$-space and time of the structure factor.
A representative scattering pattern for the simulations conducted here
is presented in Fig.~1 as a log-scale density plot. Because the
simulation samples the order parameter at $L^d$ locations in real
space, the scattering function can only be sampled at $L^d$
independent locations in $\bf k$-space. This corresponds in a natural way
to the independent speckles seen in experiments.

Based on the scaling relations in Eqs.~(\ref{eq:scalq})
and~(\ref{eq:scalt}) one might naively expect that the position of an
individual speckle should move as $k \propto t^{-n}$ as the average
structure factor changes according to the scaling form. This is not
the case.  Instead, individual speckles are observed to brighten and
dim with significant fluctuations around the time-dependent average
intensity.  The evolution of the simulated speckle intensities for a
cut along the $k_x$ axis for a single simulation is presented in
Fig.~2.  They bear a striking similarity to recent experimental data
for Cu$_3$Au \cite{DufresneThesis}. From this figure it is clear that
as the system evolves, individual speckles do {\em not} travel through
$\bf k$-space. In the simulations reported here, the correlations between
neighboring speckles were found to be on the order of $1\%$, so each
speckle could be treated as independent.

In this paper the time-time correlation in the scattering intensity at
fixed ${\bf k}$ is obtained numerically via the time-dependent
Ginzburg-Landau model for a non-conserved order parameter. The details
of this approach are presented in the next section. Section III
presents the results. The scattering intensity is shown to be an
exponentially distributed random variable. The simulation results are
also used to test the scaling ansatz for the two-time intensity
correlation. The results here suggest that the intensity correlation
time grows as $kt^{1/2}.$ Section IV presents a review of this work
along with some future directions we intend to pursue.

\section{Method}

We performed a standard simulation of the dynamics of phase ordering 
following a quench through an order-disorder phase transition 
in a system described by a nonconserved scalar order-parameter 
field $\psi({\bf x},t)$. 
We used  the time-dependent Ginzburg-Landau equation, 
\begin{equation}
\frac{\partial \psi({\bf x},t)}{\partial t} =
- \frac{\delta {\cal F}[\psi({\bf x},t)]}{\delta \psi({\bf x},t)}
+ \zeta({\bf x},t) \;,
\label{eq:TDGL}
\end{equation}
which we here have expressed in dimensionless variables for convenience. 
The first term in Eq.~(\ref{eq:TDGL}) 
corresponds to deterministic relaxation towards a
minimum value of the free energy ${\cal F}[\psi({\bf x},t)]$, and
the second term represents thermal noise. Since the late-time dynamics
are controlled by a zero-temperature fixed point, we ignore the noise
term, using only the deterministic part of Eq.~(\ref{eq:TDGL}).

We employed the standard Ginzburg-Landau-Wilson $\psi^4$ free energy
\cite{Hohenberg:77} with unit parameters. 
The deterministic dynamical equation then becomes 
\begin{equation}
\frac{\partial\psi({\bf x},t)}{\partial t}
  = \left( 1 + \nabla^2 \right) \psi({\bf x},t) 
  - \psi^3({\bf x},t) \;.
\label{eq:EOM} 
\end{equation}
Note that by choosing the customary parameters in the free energy and
the dynamical equation equal to unity, we have merely specified the
length, concetration, and time scales in the simulation.
Equation~(\ref{eq:EOM}) can therefore be used to represent the
low-temperature dynamics of Model A without loss of generality. Since
thermal fluctuations are explicitly excluded, the randomness they
introduce into the domain pattern at early
times\cite{Hernandez-Garcia:92} is implemented by an initial condition
where $\psi({\bf x},0)$ consists of independent Gaussian random
numbers with mean $0$ and standard deviation $0.1$.  The simulations
were conducted on square lattices with periodic boundary conditions,
lattice constant $\Delta x=1\,$, and a system size of
$L_x=L_y=L=1024$.  The Laplacian in Eq.~(\ref{eq:EOM}) was
approximated by the standard four-neighbor discretization, and a
simple Euler integration scheme with $\Delta t=0.05$ was used to
collect data every $20$ time units up to a maximum of $t=1000$.
Results were averaged over 10 independent initial conditions.

The Fourier transform $\hat\psi$ of $\psi$ is defined as 
\begin{equation} 
\hat\psi({\bf k},t) = \frac{1}{\sqrt{L^d}} \sum_{\bf x} \psi({\bf x},t)
                      e^{i{\bf k\cdot x}} \;,
\label{eq:FT} 
\end{equation}
where the fact that the lattice spacing is unity in all directions has
been used. The Brillouin zone is
defined by the discrete set of wavevectors, $k_x$,~$k_y = 2 \pi j/L$
with $j \in \{ 0, \, \pm1, \, \pm2, \, \dots \, \pm (L/2 -1), \, L/2
\}$.  The structure factor of the system is
\begin{equation}
S({\bf k},t)=|\hat\psi({\bf k},t)|^2
\end{equation}
and is proportional to the scattering intensity observed experimentally. 
To be consistent with the
numerical integration, the magnitude of the wavevector, $k({\bf k})$, is
defined using the operator relation
\begin{equation}
-\left[k({\bf k})\right]^2 \hat\psi({\bf k},t) = 
\frac{1}{\sqrt{L^d}} 
\sum_{\bf x} e^{i{\bf k \cdot x}} \nabla^2 \psi({\bf x},t) \;.
\end{equation}
Substituting the discrete version of the Laplacian, 
one gets  
\begin{equation}
k({\bf k})= \left[2 \left( d-\sum_{\alpha=1}^d \cos\left(k_\alpha \right) 
\right) \right]^{1/2} . 
\label{eq:knorm} 
\end{equation}
This has been used for the two-time correlation function discussed in
the next section where the value $k$ is used to bin data during
circular averaging and to determine the value of scaling
variables. Because of lattice effects we consider only those
wavevectors with $k<1$, for which the finite interface thickness can be 
ignored in this work. 

\section{Results}

The measured intensity at a given $({\bf k},t)$ is a random variable
with characteristics determined by the scattering system.  As long as
the independent 
domains are randomly located and oriented, and the domain size is
much smaller than the system size, one may expect the central limit
theorem to imply that the Fourier transformed order parameter,
$\hat{\psi}({\bf k},t)$, is well approximated by a complex Gaussian
random field, even if the real-space order-parameter field, $\psi({\bf
x},t)$, is not Gaussian. One consequence of this is that the
probability density function for the scattering intensity
$S({\bf k},t)$ can be shown to be exponential for all $\bf k$ and $t$
\cite{DufresneThesis,Wilson:49,BrownEtAl}.  The probability density
for the {\it normalized} intensities, $s({\bf k},t)={S({\bf
k},t)}/{\langle S({\bf k},t)\rangle}$, is therefore
\begin{equation}
P(s) = \exp{(-s)} \;,
\label{eq:exppdf}
\end{equation}
independent of $\bf k$ and $t$. The probability density $P(s)$ is
normalized and has unit mean and standard deviation.  Since $P(s)$ is
identical for all values of ${\bf k}$ and $t$, only one density
function needs to be constructed. This leads to very good
statistics because on the order of $10^7$ samples are available. The
results for all ${\bf k}$ and $t$, with $\langle S({\bf k},t)\rangle$
found by averaging over wavevectors with equivalent values of
$k_x^2+k_y^2$, are presented in log-linear form in Fig.~3. The solid
line is the expected density $P(s)=\exp(-s)$. For the simulation the
probability density is exponential for all $x \alt 12$, although
deviations may occur above this. We have verified that the
distribution at given values of ${\bf k}$ and $t$ also is exponential.

One important consequence of an exponentially distributed intensity is
that the average scattered intensity at $({\bf k},t)$ is equal to the
standard deviation of that intensity. This property can be used to
simplify the normalized correlation function
\begin{eqnarray}
\label{eq:corr}
{\rm Corr}({\bf k},t_1,t_2) & = & 
\frac
{\Big\langle S({\bf k},t_1)S({\bf k},t_2)\Big\rangle
-\Big\langle S({\bf k},t_1)\Big\rangle\Big\langle S({\bf k},t_2)\Big\rangle}
{\sqrt{\Big\langle S^2({\bf k},t_1)\Big\rangle-\Big\langle S({\bf k},t_1)\Big\rangle^2}
 \sqrt{\Big\langle S^2({\bf k},t_2)\Big\rangle-\Big\langle S({\bf k},t_2)\Big\rangle^2}} \\
& = &
\frac{\Big\langle S({\bf k},t_1)S({\bf k},t_2)\Big\rangle
-\Big\langle S({\bf k},t_1)\Big\rangle\Big\langle S({\bf k},t_2)\Big\rangle}
{\Big\langle S({\bf k},t_1)\Big\rangle\Big\langle S({\bf
k},t_2)\Big\rangle} \;,
\end{eqnarray}
which has a value between $-1$ and $1$ by construction. At $t_1=t_2,$
the same-time limit, the exponential nature of $s$ makes the
correlation unity as one would expect. As the two measurement times
become widely separated, the values of the intensity become
independent and the correlation decays to zero. For this relaxational
system negative correlations are not expected. The scaling
ansatz extended to this situation allows comparison of the scattering
at different ${\bf k}$ and $t$ by
\begin{equation}
{\rm Corr}(\theta_1,\theta_2) = 
\frac{\Big\langle F(\theta_1) F(\theta_2) \Big \rangle}
     {\Big\langle F(\theta_1) \Big \rangle 
     \Big \langle F(\theta_2)\Big\rangle } - 1 \;,
\end{equation}
where $\theta_{1,2}=2k^2t_{1,2}$ are the rescaled times from
Eq.~(\ref{eq:scalt}). The ansatz has been tested with the simulation
data. Because of the amount of data involved, it was necessary to bin
the results. The results for $\langle F(\theta)\rangle$ were binned by
$t$ and $\theta.$ For each pair of times, the value of
$\langle F(\theta_1)F(\theta_2)\rangle$ for a single ${\bf k}$ was
found and then placed into a bin in the
$(\theta_1,\theta_2)$~plane. The average result for each bin was later
found.  The loss of precision from this procedure should not
significantly affect the results. This function is most easily viewed
as a contour plot in the $(\theta_1,\theta_2)$~plane, as in Fig.~4. In
this figure a ${\rm Corr}(\theta_1,\theta_2)=1$ contour, which is not
shown, extends along the diagonal. Moving away from that line,
contours at values of $0.5$, $0.1$, and $0.025$ are shown. Scatter in
the data becomes dominant for values less than this. The striking
feature of this figure is that the correlation function becomes
``broader'' for later times, {\it i.e.}  the correlation time for the 
speckle intensity becomes longer as the phase ordering proceeds.

An alternative set of variables for this problem are
$\tau=(\theta_1+\theta_2)/2$ and $\delta=\theta_2-\theta_1.$ A constant
value of $\tau$ corresponds to a line perpendicular to the
$\theta_1=\theta_2$ diagonal, while $\delta$ measures the distance (in
units of scaled time) away from the diagonal. For a given value of
$\tau$, one can measure the characteristic time difference, 
$\delta_c,$ required  
for the correlation to decay to a value of $1/2.$ The values of
$\delta_c$ estimated by linear interpolation from the simulation
results are presented in log-log form in Fig.~5. For large values of
$\tau,$ the relationship appears to be a power law. The solid line
represents a least-squares fit which gives an exponent of $0.507 \pm
0.002$. Analytic arguments indicating that the exponent is exactly
$1/2$ will be presented elsewhere \cite{BrownEtAl}. Furthermore, at
these $\tau$ the correlation as a function of $\delta$ can be
collapsed onto a single master curve using $\delta_c.$ The results for
a wide range of $\tau$ are presented in Fig.~6, where the interpolated
value of $\delta_c$ has been used. The collapse of the data is quite
good, indicating that at large $\tau$ the correlations can indeed be
described by a single curve.

\section{Conclusions}
\label{sec:sum}

Intensity fluctuations in the scattering of coherent light from a
two-dimensional system undergoing an order-disorder transition have
been investigated by numerical simulation. The ratio of the measured
intensity at given $({\bf k},t)$ to its mean value at the same $({\bf
k},t)$ is found to be an exponentially distributed random number for
the dynamically evolving system. The two-time intensity correlation
function is found to obey scaling. At late times, its shape in the
direction perpendicular to the $t_1=t_2$ line, in units of
$k^2(t_2-t_1)$, is constant except for a characteristic time difference 
which increases as the system evolves. This characteristic decay time
is found to scale as $k(t_1+t_2)^{1/2}.$

Recently we have found that when the domain size is much smaller than
the system size, the speckle correlation function defined in
Eq.~(\ref{eq:corr}) can be related to the two-time correlation
function of the order parameter in real space. Theories for this
latter function exist \cite{Yeung:90,liu:91b}. We are currently
analyzing these theories in terms of the predictions made above and
investigating how coherent X-ray scattering could be used to test
these theories \cite{BrownEtAl}.

\section{Acknowledgments}
We appreciate useful discussions with E.~Dufresne, M.~Sutton, and
B.~Morin.  P.~A.~R.\ is grateful for hospitality and support at McGill
and Kyoto Universities. This research was supported by the Natural
Sciences and Engineering Research Council of Canada, {\it les Fonds
pour la Formation de Chercheurs et l'Aide \`a la Recherche de
Qu\'ebec\/}, and by the Florida State University (FSU) Center for
Materials Research and Technology, the FSU Supercomputer Computations
Research Institute under U.S.\ Department of Energy Contract No.\
DE-FC05-85ER25000, and by U.S.\ National Science Foundation grants
No.\ DMR-9315969 and INT-9512679.


\newpage

\section*{Figure Captions}

\begin{description}

\item {Fig.~1} Example of the speckled scattering intensity from one
measurement of the simulation. This is a log-scale intensity plot,
with lighter shades indicating brighter speckles. The ${\bf k}=0$
origin is at the center of the figure. Speckles do not shift in
$\bf k$-space, but their intensities fluctuate strongly around the $({\bf
k},t)$ dependent average value.

\item {Fig.~2} Example of the evolution of certain speckles with
time. This is a cut along the $k_y=0$ axis for one simulation; the
intensity scale matches that for Fig.~1. The earliest time is at the
bottom of the figure and time increases to the latest simulation time
at the top. $k_x=0$ is the left hand column and $k_x$ increases towards
the right. From this figure it is clear that individual speckles do
not move in $\bf k$-space and that neighboring speckles are relatively
uncorrelated. The fluctuations in the speckle intensity are more
apparent at higher wavevectors.

\item {Fig.~3} The probability density for the normalized speckle
intensity, $s=S({\bf k},t)/\langle S({\bf k},t)\rangle$. The solid
line is the theoretical density of an exponentially distributed random
variable.

\item {Fig.~4} Contour plot of the scaled two-time correlation
function. The correlation at $\theta_1=\theta_2$ is unity by
construction and is not shown. The contours moving away from this
diagonal are at $0.5$, $0.1$, and $0.25$. The figure shows that the
speckle intensity stays correlated longer as $\theta_1+\theta_2$
increases.

\item {Fig.~5} The characteristic decay time difference, $\delta_c,$ 
as a function of $\tau \propto t_1+t_2.$ For larger values of $\tau,$
the relationship is seen to be a power law. The solid line is a
least-squares fit which finds $\delta_c \sim \tau^{0.507\pm0.002}.$

\item {Fig.~6} Collapse of the intensity-intensity correlation for
different values of $\tau \propto t_1+t_2,$ where $\delta \propto
t_2-t_1$ and $\delta_c$ is shown in Fig.~5. The figure shows that the
correlation can be collapsed onto a single curve for a large range of
$\tau.$ 

\end{description}

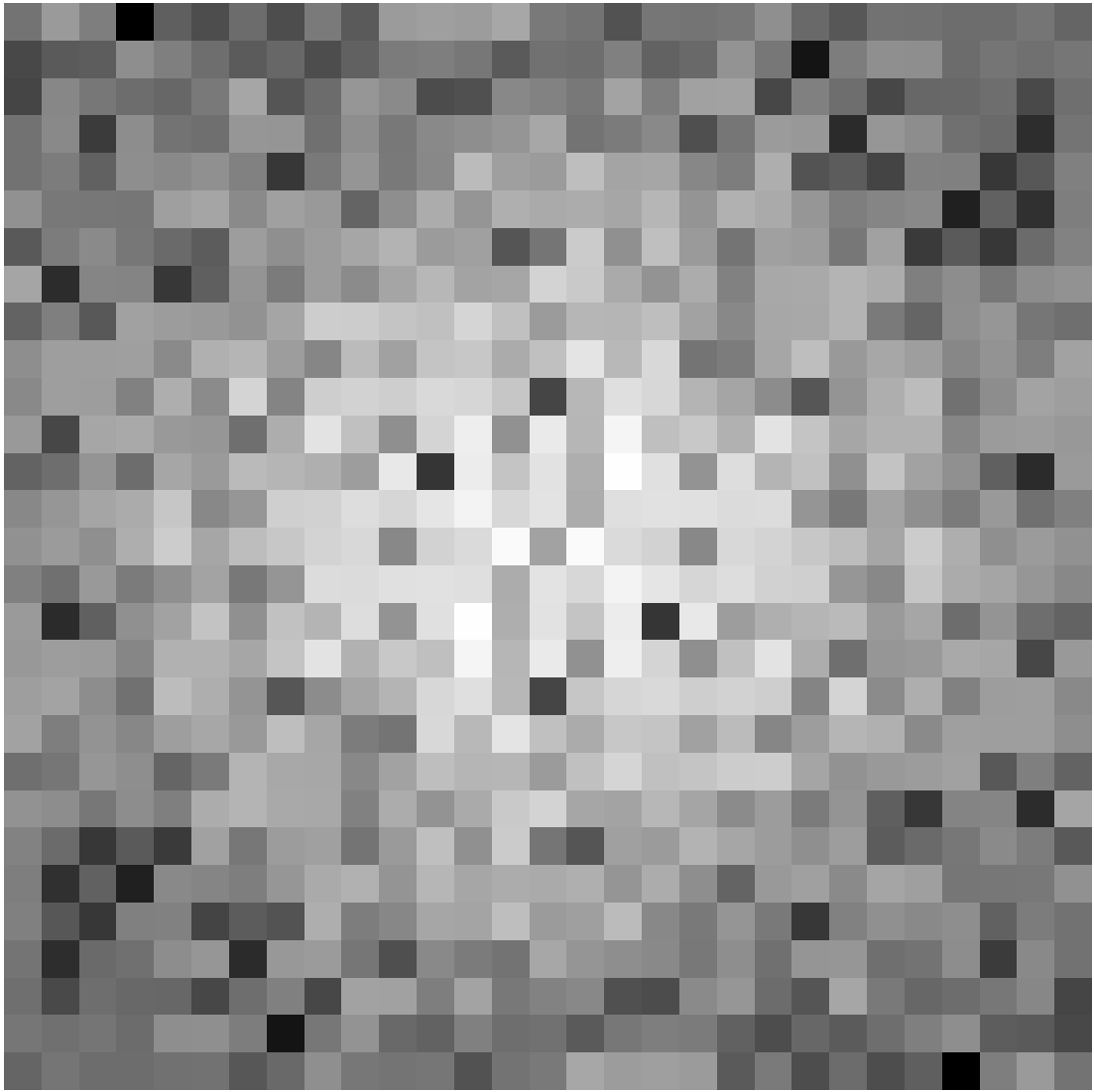
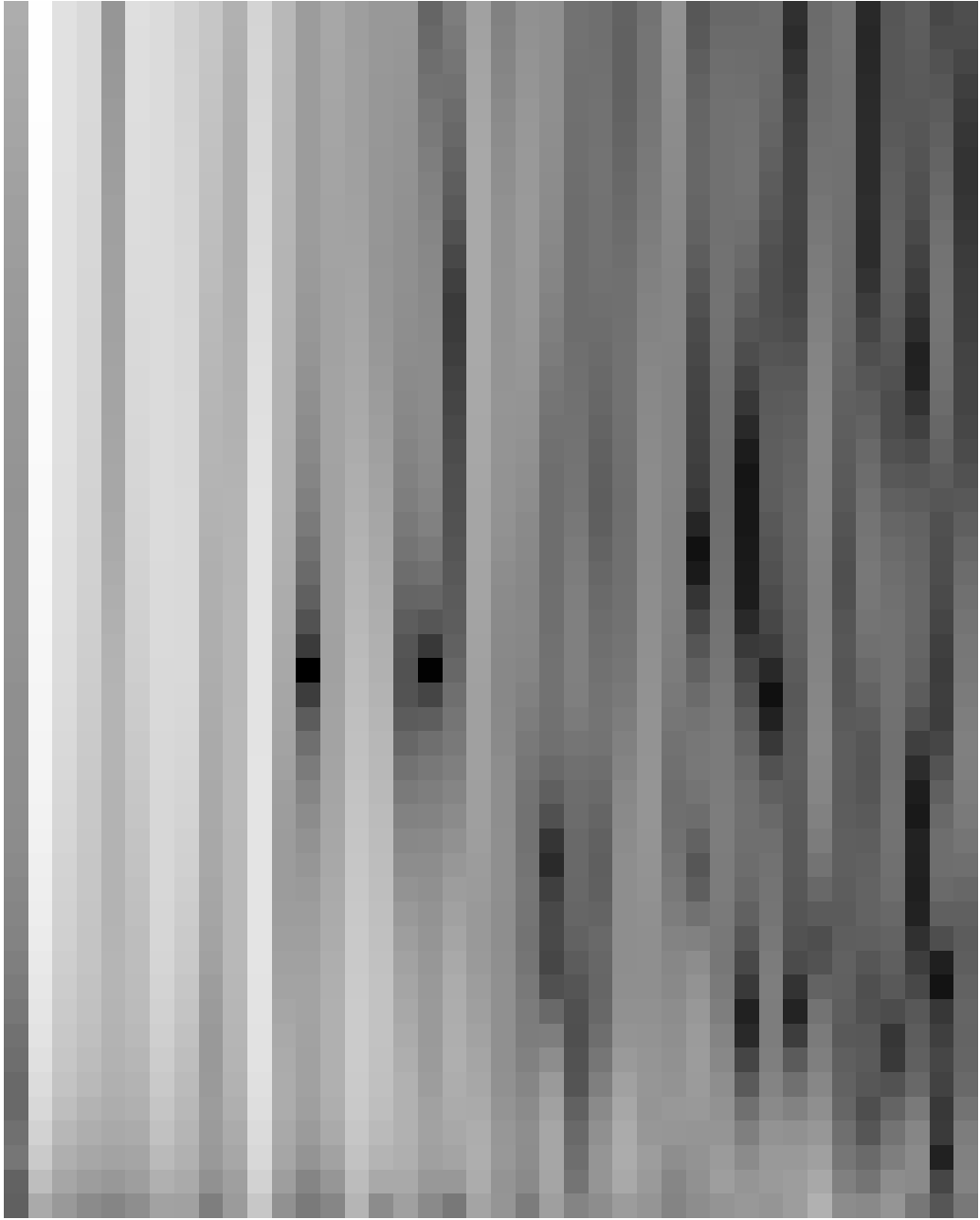
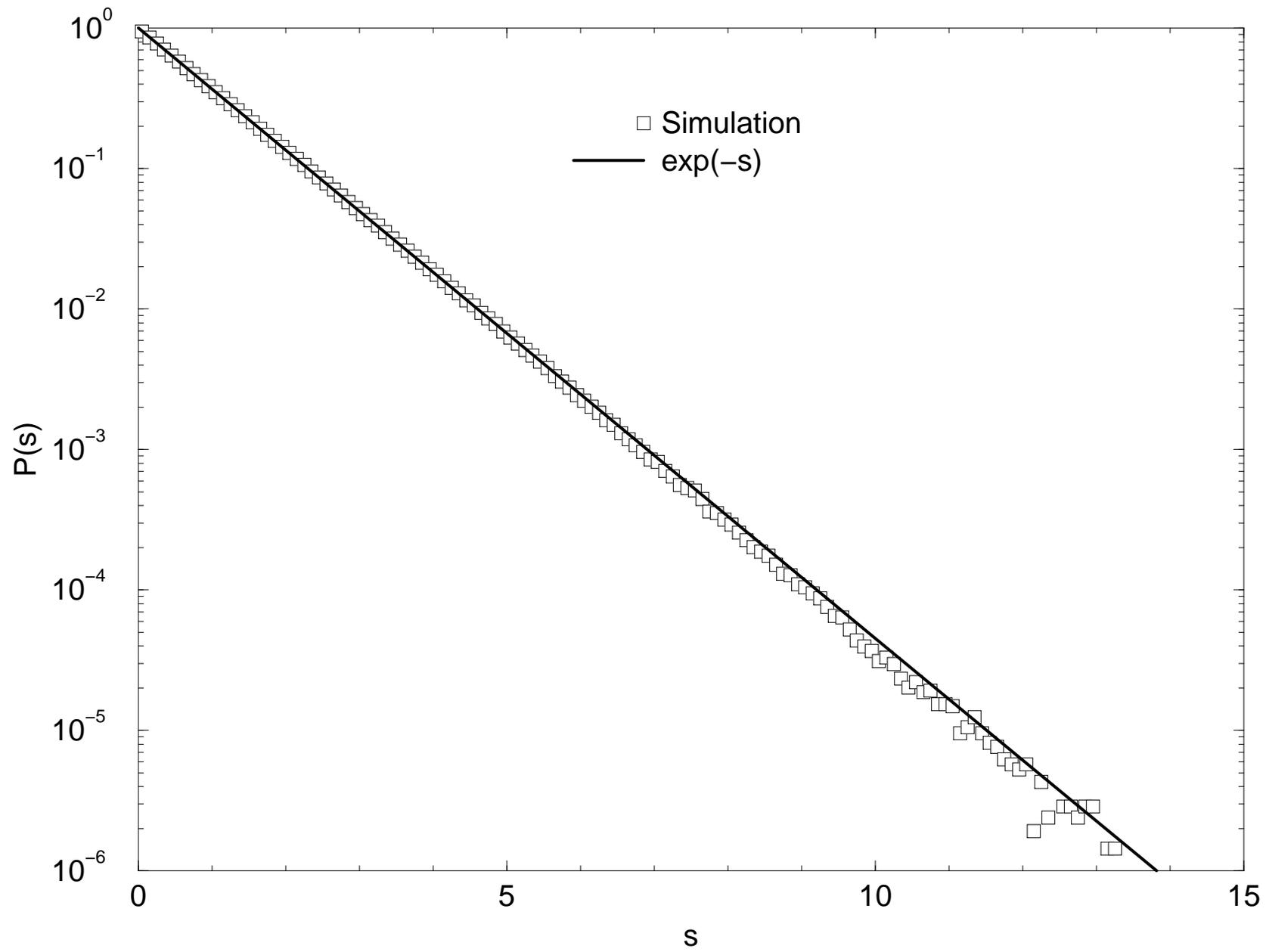
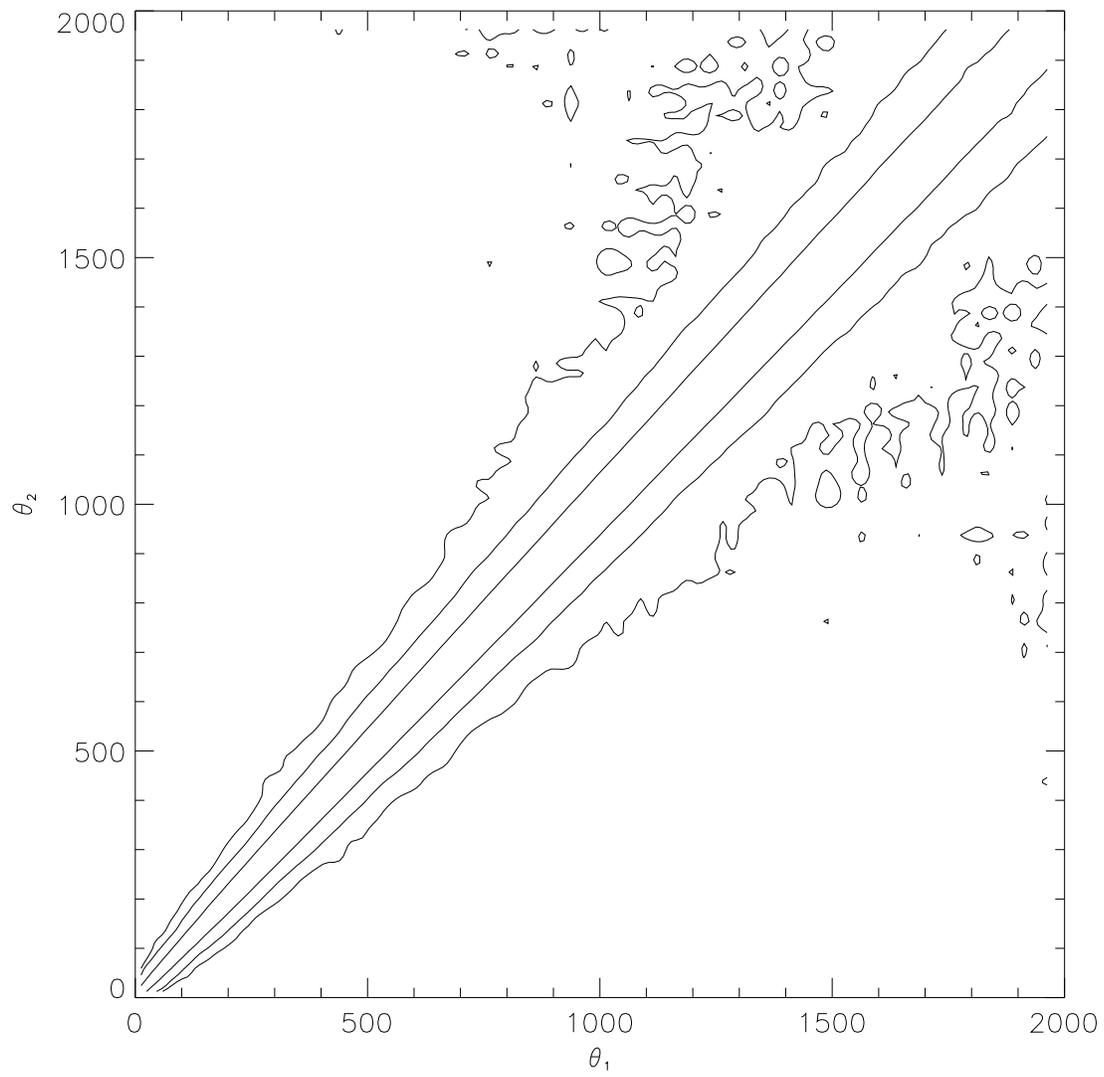
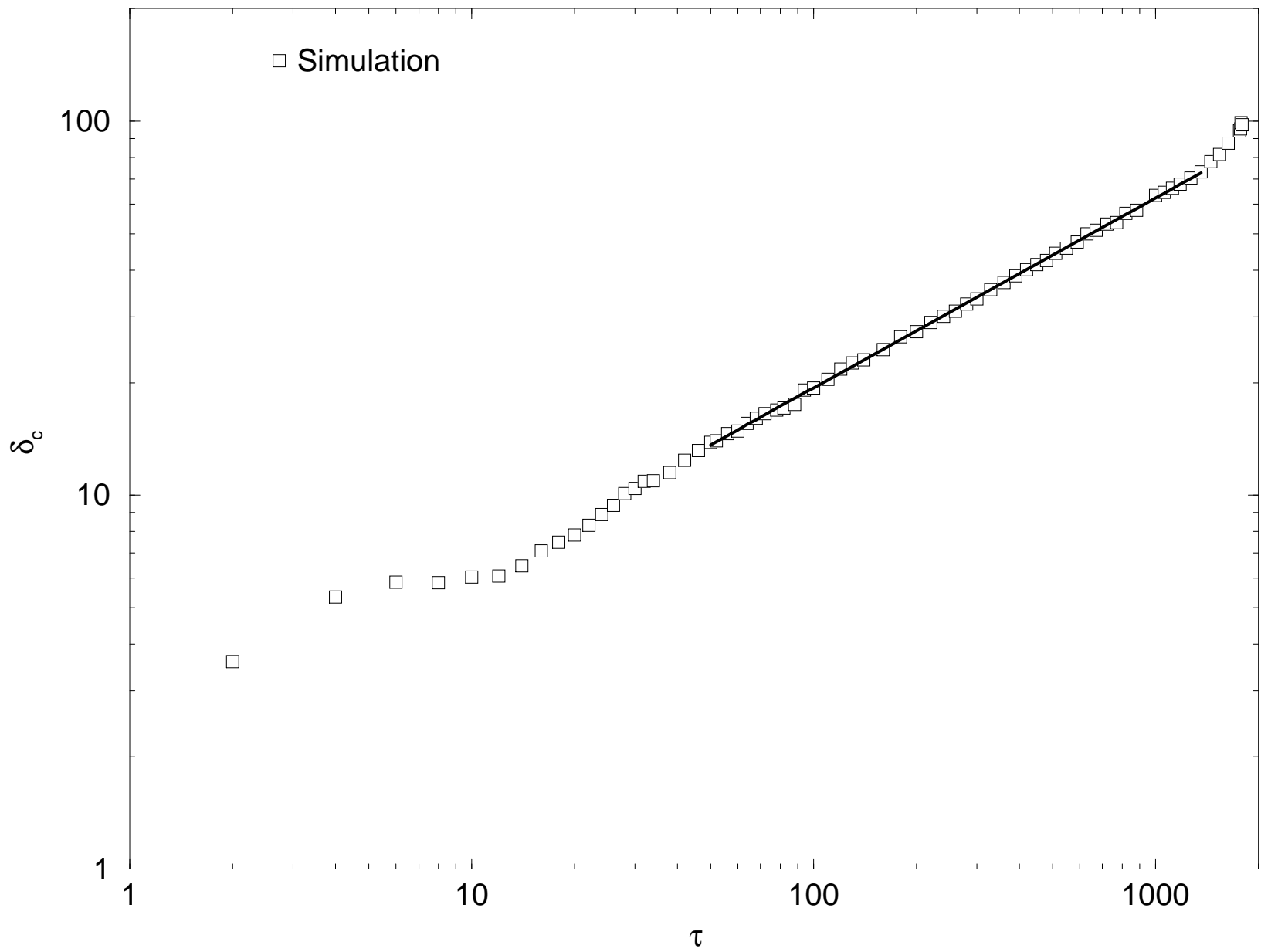
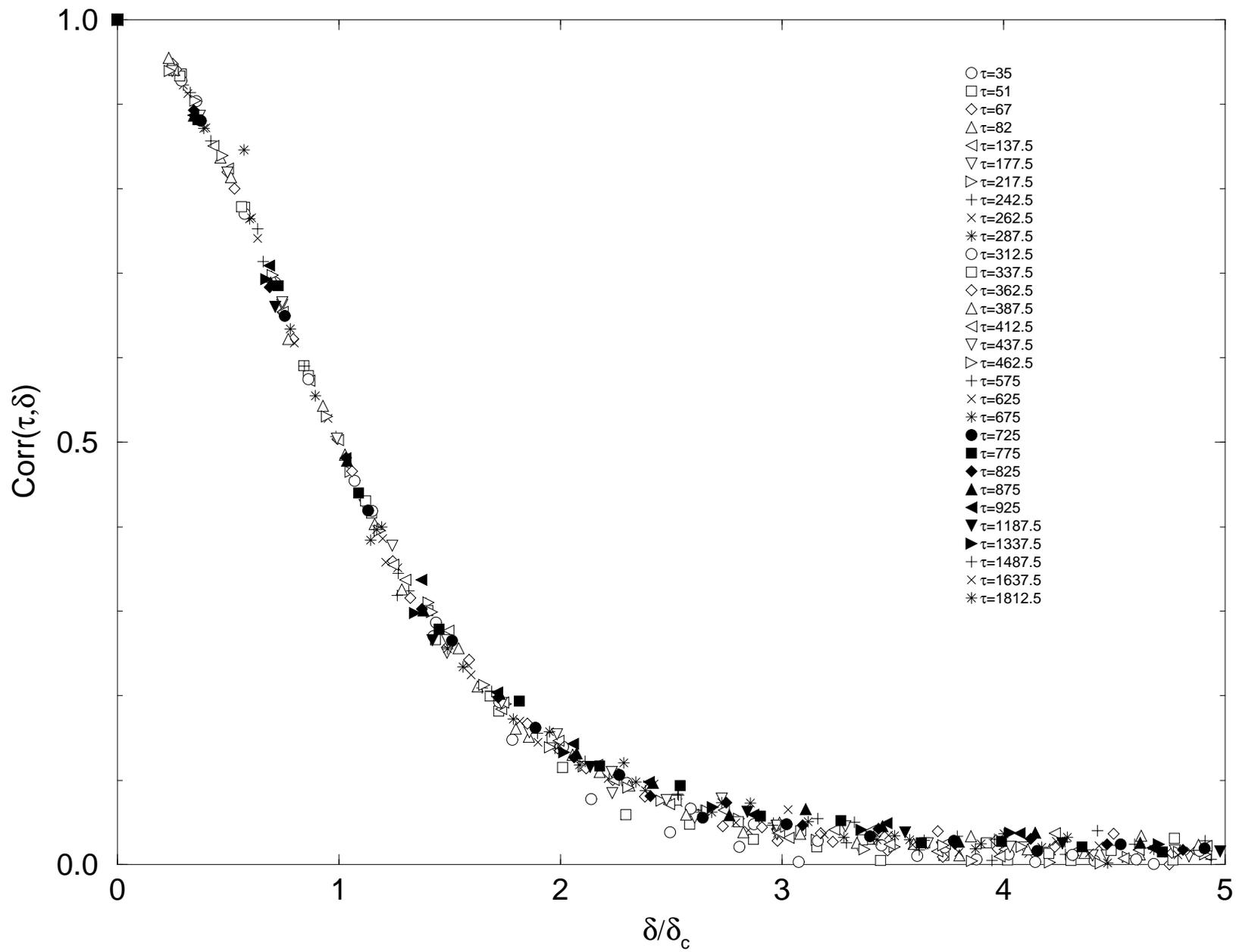
\end{document}